\documentstyle[onecolumn]{mn}

\newif\ifAMStwofonts

\ifoldfss
  \ifCUPmtlplainloaded \else
    \NewTextAlphabet{textbfit} {cmbxti10} {}
    \NewTextAlphabet{textbfss} {cmssbx10} {}
    \NewMathAlphabet{mathbfit} {cmbxti10} {} 
    \NewMathAlphabet{mathbfss} {cmssbx10} {} 
  \fi
  \ifAMStwofonts
    \ifCUPmtlplainloaded \else
      \NewSymbolFont{upmath} {eurm10}
      \NewSymbolFont{AMSa} {msam10}
      \NewMathSymbol{\upi}     {0}{upmath}{19}
      \NewMathSymbol{\umu}     {0}{upmath}{16}
      \NewMathSymbol{\upartial}{0}{upmath}{40}
      \NewMathSymbol{\leqslant}{3}{AMSa}{36}
      \NewMathSymbol{\geqslant}{3}{AMSa}{3E}

      \let\leq=\leqslant 
      \let\geq=\geqslant 
    \fi
  \fi
\fi 

\ifnfssone
  \newmathalphabet{\mathit}
  \addtoversion{normal}{\mathit}{cmr}{m}{it}
  \addtoversion{bold}{\mathit}{cmr}{bx}{it}
  \newmathalphabet{\mathbfit} 
  \addtoversion{normal}{\mathbfit}{cmr}{bx}{it}
  \addtoversion{bold}{\mathbfit}{cmr}{bx}{it}
  \newmathalphabet{\mathbfss} 
  \addtoversion{normal}{\mathbfss}{cmss}{bx}{n}
  \addtoversion{bold}{\mathbfss}{cmss}{bx}{n}
  \ifAMStwofonts
    \ifCUPmtlplainloaded \else
      \UseAMStwoboldmath
      \makeatletter
      \new@mathgroup\upmath@group
      \define@mathgroup\mv@normal\upmath@group{eur}{m}{n}
      \define@mathgroup\mv@bold\upmath@group{eur}{b}{n}
      \edef\UPM{\hexnumber\upmath@group}
      \new@mathgroup\amsa@group
      \define@mathgroup\mv@normal\amsa@group{msa}{m}{n}
      \define@mathgroup\mv@bold\amsa@group{msa}{m}{n}
      \edef\AMSa{\hexnumber\amsa@group}
      \makeatother
      \mathchardef\upi="0\UPM19
      \mathchardef\umu="0\UPM16
      \mathchardef\upartial="0\UPM40
      \mathchardef\leqslant="3\AMSa36
      \mathchardef\geqslant="3\AMSa3E

      \let\leq=\leqslant 
      \let\geq=\geqslant 
    \fi
  \fi
\fi 

\ifnfsstwo
  \DeclareMathAlphabet{\mathbfit}{OT1}{cmr}{bx}{it}
  \SetMathAlphabet\mathbfit{bold}{OT1}{cmr}{bx}{it}
  \DeclareMathAlphabet{\mathbfss}{OT1}{cmss}{bx}{n}
  \SetMathAlphabet\mathbfss{bold}{OT1}{cmss}{bx}{n}
  \ifAMStwofonts
    \ifCUPmtlplainloaded \else
      \DeclareSymbolFont{UPM}{U}{eur}{m}{n}
      \SetSymbolFont{UPM}{bold}{U}{eur}{b}{n}
      \DeclareSymbolFont{AMSa}{U}{msa}{m}{n}
      \DeclareMathSymbol{\upi}{0}{UPM}{"19}
      \DeclareMathSymbol{\umu}{0}{UPM}{"16}
      \DeclareMathSymbol{\upartial}{0}{UPM}{"40}
      \DeclareMathSymbol{\leqslant}{3}{AMSa}{"36}
      \DeclareMathSymbol{\geqslant}{3}{AMSa}{"3E}

      \let\leq=\leqslant 
      \let\geq=\geqslant 
    \fi
  \fi
\fi 

\ifCUPmtlplainloaded \else
  \ifAMStwofonts \else 
    \def\upi{\pi}
    \def\umu{\mu}
    \def\upartial{\partial}
  \fi
\fi

\newcommand{\beq}{\begin{equation}}
\newcommand{\beqa}{\begin{eqnarray}}
\newcommand{\eeq}{\end{equation}}
\newcommand{\eeqa}{\end{eqnarray}}
\newcommand{\siml}{\la}
\newcommand{\simg}{\ga}

\journal{YITP-98-34, KUNS-1515}

\title{
Observational tests of x-matter models
}
\author[T.Chiba et al.]{Takeshi Chiba$^1$, Naoshi Sugiyama$^2$ and 
Takashi Nakamura$^3$\\
$^1$Department of Physics, University of Tokyo, Tokyo 113-0033, Japan\\
$^2$Department of Physics, Kyoto University, 
Kyoto 606-8502, Japan\\
$^3$Yukawa Institute for Theoretical Physics, Kyoto University, 
Kyoto 606-8502, Japan
}

\date{Accepted .
      Received ;
      in original form}

\pagerange{\pageref{firstpage}--\pageref{lastpage}}
\pubyear{1998}

\begin{document}

\maketitle

\label{firstpage}

\begin{abstract}
  We study gravitational lensing statistics, matter power spectra and
  the angular power spectra of the cosmic microwave background (CMB)
  radiation in x-matter models.  We adopt an  equation-of-state of
  x-matter which can express a wide range of matter from the
  pressureless dust to the cosmological constant. A new ingredient in this
  model is the sound speed of the x-component in addition to the
  equation-of-state $w_0=p_{\rm x0}/\rho_{\rm x0}$. Except for the
  cosmological constant case, the perturbations of x-matter itself are 
  considered. Our primary interest is in the effect of non-zero sound
  speed on the structure formation and the CMB spectra. 
  It is found that there exist
  parameter ranges where x-matter models are consistent with all
  current observations. The x-matter generally  leaves imprints in the
  CMB anisotropy and the matter power spectrum, which should be
  detectable in  future observations.

\end{abstract}


\section{introduction}

There is growing observational evidence that the mean mass density
of the universe is less than the critical density. 
The growth of large scale structure in cold dark matter(CDM) models
requires $\Gamma=\Omega_{M}h=0.25\pm 0.05$ \cite{pd} where $\Omega_M$ is
the ratio of the matter energy density to the critical density and $h$ 
is the Hubble parameter in units of $100 {\rm km s}^{-1}{\rm Mpc}^{-1}$. 
Recent observations of high-redshift Type Ia
supernovae  also indicate $\Omega_{M} < 1$ \cite{sn1,sn2,sn3,sn4}.
The observations of the 
age of the globular clusters $12\pm 2{\rm Gyr}$ \cite{gc}
and the Hubble parameter $H_0 = 65\pm
10$km/s/Mpc \cite{hubble1,hubble2}, though there might be 
some modifications by recent Hipparcos results \cite{fc,reid}, 
indicate $H_{0} t_0 \simg 0.8$ where $t_0$
is the age of the universe.
All these observational pieces of evidence may not be explained in the
framework of standard CDM model(sCDM) in which $\Omega_{M}=1$.  
The non-zero cosmological constant ($\Lambda$CDM)may be needed to increase
the age of the universe as well as to make  the total energy 
density equal to the critical density as generically predicted by
inflationary cosmology  \cite{os}.
However, as  noted in Nakamura et al.\shortcite{nncs}, 
at least for the age problem,  the cosmological constant 
is  not the only way to solve the problem: if $H_0t_0 > 1$, then 
what is needed in general is the  ``matter'' which violates the 
strong energy condition such that $\rho+3p < 0$.

 Turner and White \shortcite{tw} proposed xCDM 
as an alternative to $\Lambda$CDM models in which a
 nonclustering matter of $p_{\rm x}=w\rho_{\rm x}$ 
($w$ being constant with $-1 < w  < 0$) resides in addition to
ordinary CDM so that the universe has a flat geometry. They also noted 
some  topological defects behave like  the ``x-component''.  
However, as they admit, a component with $w<0$ is highly unstable 
to growth of perturbations on small scales. 
Therefore they assumed xCDM to be a smooth component on small scales. 
The notion of a smooth component is ill-defined 
because i) the smoothness is not gauge-invariant and ii) 
it is unphysical to ignore the response of the
component to the inhomogeneities in other cosmological fluid
under the equivalence principle \cite{qmatter}.

Recently, we have constructed a phenomenological energy momentum tensor
for ``x-matter''\cite{csn}which i) is locally stable, and ii) obeys the
causality.  To satisfy the local stability requirement, we have to
separate the sound speed $c_s^2\equiv \dot{p_{\rm x}}/\dot{\rho_{\rm 
x}}$ from the coefficient of the equation of state $w_{\rm x} \equiv
p_{\rm x}/\rho_{\rm x}$.  We assume x-matter obeys the equation of state
such that $p_{\rm x}=w_0\rho_{\rm x0} + c_s^2(\rho_{\rm x}-\rho_{\rm
  x0})$, where $p_{\rm x}$, $\rho_{\rm x}$ and $\rho_{\rm x0}$ are the
pressure, the energy density and the present energy density of x-matter,
respectively.  As regards the background universe, this x-matter model
contains both Einstein-de Sitter ($w_0=c_{s}^2=0$) 
and $\Lambda$ universes ($w_0=-1$) as a limit.
A new ingredient in this
model is the sound speed of the x-component in addition to the
equation-of-state $w_0$.  Since at present we do not know how the vacuum
energy interacts gravitationally \cite{weinberg}, a phenomenological
approach taken here is worthwhile.

On the other hand, when we deal with perturbations of x-matter to study
the structure formation in the universe, the limit of the $\Lambda$
universe in which density and pressure stay constant has to be carefully
taken.  In this case, constant density and pressure allow the existence
of only smoothed distribution from the gauge-invariance.  Therefore
x-matter is equivalent to the cosmological term.

In this paper, we shall make further observational tests of x-matter
models as alternatives to sCDM or $\Lambda$CDM.  First we consider the
statistics of gravitational lensing of quasars.  Then we study the
evolution of density perturbations of x-matter models.  Accordingly, we
obtain observational quantities such as the matter power spectrum and
cosmic microwave background (CMB) anisotropies. The class of
cosmological models we consider are spatially flat
Friedmann-Robertson-Walker(FRW) models which are generically predicted
by cosmological inflation.  We consider x-matter, dust matter (baryon
and/or CDM) and radiation (photon and neutrino) while we do not
include the cosmological constant term whose density is smoothly
distributed.  Similar analysis was made in xCDM models \cite{waga,waga2},
in scaling field model \cite{fj1,fj2}, the decaying lambda model \cite{vl}
 and in quintessence(QCDM) model \cite{qmatter}.
When our paper was at the nearly end of 
completion, we became aware of the recent paper by Hu \shortcite{hu}
where similar models were considered.  In his paper, he employed very
general equation of states which include our model.  However, since we
focus on the simple, yet rather general equation of state,
we could perform thorough investigation and observational tests
of the model here.

The lensing frequencies are expected to be 
reduced in x-matter models, since in a x-matter model the 
distance to an object at redshift $z$ is smaller than the distance 
to the same object in  $\Lambda$-dominated models.
We study the lensing frequencies and set
quantitative limits on  the present energy density of x-matter using 
quasar surveys.

The height and location of the peak in the angular power spectrum of
CMB depend on the matter-radiation equality and 
the expansion rate of the universe at the surface of last scatter. 
Different equation-of-states yields the different matter-radiation
equality time,
which would leave imprint in the peak height.
Non-zero $c_s$ affects the expansion rate of the universe, and its
effect  would leave the imprint in the location of the peak even if
for low $\Omega_{\rm x}$. 

This paper is organized as follows. In \S 2, we review our x-matter
model. In \S 3, we present calculations of gravitational lensing in
this model. In \S 4, a detailed perturbation theory in x-matter models 
is given and the spectra of cosmic microwave background are calculated. 
\S 5 is devoted to summary. 
We use the units of $c=1$. 

\section{x-matter model}

In this section we review our x-matter model \cite{csn}. 

x-matter  possesses following nature, i.e., 
$w_{\rm x} \equiv p_{\rm x}/\rho_{\rm x} < 0$ to elongate the cosmic
age and $c_s^2 \equiv \delta p /\delta \rho \geq 0$ to stabilize
growth of perturbations. 
$c_s$ represents the sound speed of x-matter and
determines how the density evolves. 
Together with the latter condition,
we demand that it is not faster than the speed of light from the causality.  
As a simple form, we employ the equation of state of
x-matter as 
\beq p_{\rm x}=w_0\rho_{\rm x0}+c_s^2(\rho_{\rm x}-\rho_{\rm x0}), 
\label{eos}
\eeq 
where
the suffix $0$ denotes the present value.
Although $w_0$ and
$c_s$ are functions of $\rho_{\rm x}$ in general, to simplify the
discussion we assume that they are constants. 
Our x-matter model includes the cosmological constant model 
as a special case since $p_{\rm x}=-\rho_{\rm x0}$ for 
$w_0=-1$ regardless of the value of $c_s^2$.
However, as is mentioned in the previous section, 
it should be noticed that x-matter is equivalent to the cosmological term
and it is no longer `matter' in this limit.  
According to above requirements, we consider the region $-1\leq w_0 \leq 0$, 
$0 \leq c_s^2 \leq 1$, and $\rho_{\rm x} \geq 0$. 
{} From the requirement that  
the process of Big Bang Nucleosynthesis (BBN) is not appreciably 
disturbed  by the existence of x-matter, $c_s^2$ is further
constrained: $0\leq c_s^2 \siml 0.15$ \cite{csn}. 

The energy equation is 
\beq
{d \rho_{\rm x}\over d a}=-{3\over a}(\rho_{\rm x}+p_{\rm x}) ,
\eeq
where $a$ is the scale factor. Normalising $a$ as  $a_0=1$, we have 
\beq
\rho_{\rm x}={\rho_{\rm x0}\over 1+c_s^2}\left
( (1+w_0)a^{-3(1+c_s^2)}+c_s^2-w_0\right).
\label{rho}
\eeq
Note that $w_0=-1$ corresponds to the cosmological constant model
irrespective of $c_s^2$. We also note that the cosmological constant
model is an asymptotic limit in 
our x-matter model: $p_{\rm x}/\rho_{\rm x} \rightarrow -1$ as
$a\rightarrow \infty$. The effective scalar field potential for the
x-matter model is described in Appendix A.

The Friedmann equation in a flat universe is
\beq
{1\over H_0^2}\left({\dot{a}\over a}\right)^2 
={\Omega_{M}\over a^3}   
+ {\Omega_{r}\over a^4}
+{\Omega_{\rm x}\over 1+c_s^2}\left 
( (1+w_0)a^{-3(1+c_s^2)}+c_s^2-w_0\right), 
\label{frw} 
\eeq
where $\Omega_{M}$, $\Omega_r$ and $\Omega_{\rm x}$ are the present 
density parameter of matter(baryon and CDM; $\Omega_{M}=\Omega_{\rm B}
+ \Omega_{\rm CDM}$), radiation (photon and neutrino)
and x-matter, respectively and over dot denotes the 
time derivative. 

The cosmic age is given by
\beq
H_0t_0=\int_0^1
{da\over
  a}\Bigl(\Omega_{M}a^{-3}
+{\Omega_{\rm x}\over 1+c_s^2}\left
( (1+w_0)a^{-3(1+c_s^2)}+c_s^2-w_0\right)
\Bigr)^{-1/2}.
\eeq
Here we have neglected the energy density of the radiation components 
since the effect of them on the cosmic age is negligible. 
The functional form of $H_0t_0$ shows that it is a decreasing function 
of $c_s^2$ or $w_0$. In Fig.1, the contour plots for $H_0t_0$ in the
$(w_0,\Omega_{\rm x})$ plane are shown ($\Omega_M=1-\Omega_{\rm x}$).

\section{Gravitational Lensing}

Now we consider the statistics of gravitational lensing. 
We compute the predicted number of gravitational lensing events taking 
into account the magnification bias and compare it with the
observational data. Our analysis largely depends on
Kochanek \shortcite{kochanek} since our main interest is in the
comparison with $\Lambda$-dominated models. 
Accordingly, we will neglect the dust obscuration effect considered by 
Malhotta et al.\shortcite{turner} and the evolution effect of lensing 
galaxies which would weaken the constraint.

\subsection{Lensing Probability}

We consider a singular isothermal model for a lens mass distribution
mainly for simplicity. {\it Hubble Space Telescope(HST)}
observations show that E/S0 galaxies are effectively
singular \cite{tremaine}, which may support our assumption. 
The effect of a finite core radius is shown to be not so
significant \cite{kochanek} because a core radius increases the cross
section while decreases the amplification compared to a singular
isothermal model and both effects are compensated in the lensing
probability.
The cross section for lensing events for a singular isothermal sphere 
lens model is given by \cite{tog} (see Appendix B for its derivation)
\beq
\sigma=16\pi^3{v}^4\left({D_{OL}D_{LS}\over
D_{OS}}\right)^2,
\label{tog}
\eeq
where $v$ is the velocity dispersion of the dark halo of a lensing
galaxy, $D_{OL}$ is the
angular diameter distance to the lens, $D_{OS}$ is the
angular diameter distance to the source, and $D_{LS}$ is the
angular diameter distance between the lens and the source.
 
The differential probability of a lensing event is
\beqa
d\tau&=&n_0(1+z_L)^3B\sigma{dt\over dz_L}dz_L,\nonumber\\
&=&16\pi^3n_0v^4(1+z_L)^3B\left({D_{OL}D_{LS}\over 
D_{OS}}\right)^2{dt\over dz_L}dz_L,
\label{tau}
\eeqa
where $n_0$ is the present-day number density of lensing galaxies, and 
$B$ is the amplification bias factor defined below. 
The total lensing probability is obtained by integrating
Eq.(\ref{tau}) with respect to $z_L$.


The number count of galaxies is fitted by a Schechter function \cite{sch}
\beq
\phi_{g}(L)dL=\phi_{*}\left({L \over
L_{*}}\right)^{\alpha}e^{-L/L_{*}}{dL\over L_{*}}.
\eeq
Recent Autofib survey gives $\phi_{*}=0.026\pm 0.08h^3{\rm Mpc}^{-3},
\alpha=-1.09^{+0.10}_{-0.09} $\cite{ellis}. 
We only consider the contribution from E/S0 type galaxies.
The contribution from spiral galaxies is not so 
important \cite{kochanek,ft,mr} 
because the typical velocity dispersion ($\simeq 140 {\rm km/s}$) is 
small.  The comoving number density of E/S0 galaxies $n_e$ is 
$n_e=3.5\times 10^{-3}h^3{\rm Mpc}^{-3}$ \cite{ellis2}. 

We assume that the velocity dispersion of the dark halo $v$ 
is related to the luminosity by a ``Faber-Jackson relation'' of the form
$v=v_*(L/L_*)^{1/\gamma}$. For a singular isothermal model, Kochanek 
found that $v_*=225\pm 10{\rm km/s}$ and $\gamma=4.1\pm
0.9$\cite{koch2}.
Following Kochanek \shortcite{kochanek}, we do not
consider the $(3/2)^{1/2}$ correction in the velocity dispersion since 
the dynamically estimated  velocity dispersion  of dark matter 
is not different from the central velocity dispersion.


As for the distance formula, we adopt the standard angular diameter 
distance $d_A$ in a FRW universe given by 
\beq
d_AH_0={1\over 1+z}\int^z_0 du 
\Bigl[\Omega_{M}(1+u)^3
+{\Omega_{\rm x}\over
1+c_s^2}\left((1+w_0)(1+u)^{3(1+c_s^2)}+c_s^2-w_0\right)
\Bigr]^{-1/2}.
\eeq
Then the angular diameter distance between the lens and the source
 $D_{LS}$ is given by
\beq
D_{LS}H_0={1\over 1+z_S}\int^{z_S}_{z_L}
du 
\Bigl[\Omega_{M}(1+u)^3+{\Omega_{\rm x}\over
1+c_s^2}\left((1+w_0)(1+u)^{3(1+c_s^2)}+c_s^2-w_0\right)
\Bigr]^{-1/2}
\eeq


The magnification bias $B$ is an enhancement of probability that a
quasar is lensed. The bias for a quasar at redshift $z$ with apparent 
magnitude $m_Q$ is written as 
\beq
B(m_Q,z_Q)={\int dm_Q\int^{\infty}_{A_{min}}\phi_Q(m_Q+2.5\log
A,z_q)P(A)dA\over \int dm_Q\phi_Q(m_Q,z_Q)},
\eeq
where $\phi_Q(m_Q,z_Q)$ is the luminosity function of quasars
at redshift $z_Q$,  $A_{min}$ is the minimum total flux  amplification, 
and $P(A)$ is the probability distribution for a greater amplification 
$A$.

We use the Kochanek's ``best model''\cite{kochanek} as a quasar luminosity 
function. In this model, like Boyle, Shanks and Peterson's
model \cite{boyle}, the luminosity function has the form
\beq
\phi_Q \propto \left
( 10^{-\alpha_Q(m-m_0(z))}+10^{-\beta_Q(m-m_0(z))}\right)^{-1}, 
\eeq
where the bright-end slope $\alpha_Q$ and faint-end slope 
$\beta_Q$ are constants, while  
the break magnitude $m_0(z)$ evolves with redshift \cite{kochanek}
\beq
m_0(z)=\left\{\begin{array}{ll}
       m_0+(z-1.0),  &\  z < 1\\
       m_0,          &\ 1\leq z \leq  3\\
       m_0-0.7(z-3), &\ z>3.
\end{array}\right. 
\eeq
Fitting this evolution model to the quasar luminosity function
data in Hartwich and Schade \shortcite{hs} for $z>1$,  
Kochanek find that ``the best model'' has 
$\alpha_Q=1.07\pm 0.07, \beta_Q=0.27\pm 0.07$ and 
$ m_0=18.92\pm 0.16$ at $B$
magnitudes. He noted that this model is no better than Wallington and
Narayan model \cite{wn}. 

For a singular isothermal sphere, the amplitude amplification
$A(r)$ from a projected lens-source separation $r$ is 
\beq
A=2b_{cr}/r, 
\eeq
where $b_{cr}$ is the critical impact parameter given in
Eq.(\ref{crit}). Then $A \geq 2$ for a singular isothermal sphere. 
The probability of amplification is hence 
\beq
P(A)=8/A^3.
\eeq


Putting all these pieces together, 
the lensing probability is finally given by 
\beqa
\tau&=&\int_0^{z_Q}dz_L{dt\over dz_L}(1+z_L)^3\int_0^{\infty}{dL\over L_{*}} 
n_e v_*^4\left({L \over L_{*}}\right)^{\alpha}e^{-L/L_{*}}
16\pi^3\left({L\over L_*}\right)^{4/\gamma}B(m_Q,z_Q)
\left({D_{OL}D_{LS}\over
D_{OS}}\right)^2\nonumber\\
&=& F\int_0^{z_Q}dz_L{dt\over H^{-1}dz_L}(1+z_L)^3B(m_Q,z_Q)
\left({D_{OL}D_{LS}\over
H_0^{-1}D_{OS}}\right)^2.
\eeqa
Here $F$ is given by   
\beq
F=16\pi^3H_0^{-3}n_e
v_*^4\Gamma(4/\gamma+\alpha+1)=1.6^{+0.7}_{-0.5} \times10^{-2},
\eeq
where $\Gamma(x)$ is the gamma function. $F$  roughly means the  
probability that a quasar at horizon distance is lensed by a $L_*$ galaxy.

Fig.2 shows the optical depth in a flat universe 
 as a function of a source
redshift $z_s$ for $\Omega_{M}=0.1,0.3,0.5,0.7$, respectively. 
Current constraint \cite{kochanek} may be translated as
$\tau(z_s)/\tau(z_s;\Omega_{M}=1) \siml 5$. We expect that x-matter 
dominated models may survive the observational test. Next  we
estimate the allowed parameter regions  more quantitatively.

\subsection{Comparison with Observations}

We fix the parameters in the quasar luminosity function
($\alpha_Q,\beta_Q,m_0$) but allow changes of parameters in the galaxy 
luminosity function within their error bars. We then compare the
predicted expectation number of lenses with the observed one.  
Data are taken from HST snapshot survey \cite{maoz} and the NOT
survey \cite{not}. We use only quasars at $z>1$. The total number of
quasars in the sample is then 625. 
The sample contains 4 lensing events caused by
galaxies(0142-100,1115+080,1413+117,1268+1011). 
Since most of the quasars in the sample do not have a $B$-band
magnitude, we transform $m_V$ to a $B$-band magnitude using $B-V=0.2$
as suggested by Bahcall et al.\cite{bahcall}. 
Fig.3 shows the expected number of lenses for
$w_0=-1,-0.8,-0.6,-0.4,-0.2,0$ and 
$c_s^2=0,0.05,0.1,0.15$ as a function of $\Omega_{\rm x}$.
Each  curve from top to bottom corresponds 
to $w_0=-1,-0.8,-0.6,-0.4,-0.2,0$. $w_0=-1$ corresponds to the cosmological
constant model irrespective of $c_s^2$. The expected number increases
as $\Omega_{\rm x}$ increases. However, its rate decreases for increasing 
$w_0$ or $c_s^2$. 

We perform a maximum-likelihood analysis to determine the confidence
level of model parameters. 
The likelihood function is 
\beq
L=\Pi^{N_U}_{i=1}(1-p_i)\Pi_{j=1}^{N_L}p_j,
\eeq
where $N_U$ is the number of unlensed quasars, $N_L$ is the number of
lensed quasars, and $p_i$ is the probability that a quasar is lensed. 
$L$ is a function of three parameters $(c_s^2,w_0,\Omega_{\rm x})$.
We allow parameters $c_s^2$ and $w_0$ to vary in the range 
$ 0 \leq c_s^2 \leq 0.15, -1 \leq w_0 \leq 0$,
respectively.  
The logarithm of the ratio of the
likelihood to its maximum $-2\ln L/L_{max}$ is asymptotically
distributed like a $\chi^2$ distribution with the degrees of freedom of 
parameters involved \cite{ks}. Thus we can determine the confidence level. 

Likelihood contours are plotted in Fig.4 
projected onto constant $c_s$ plane  plane. 
Likelihood levels are, from left top
to right down, $68.3\%, 95\%, 99\%$. 
We find that larger  $\Omega_{\rm x}$ is allowed as  
increasing $w_0$. Changing $c_s$ has little effect on likelihood
contours because the effect of $c_s^2$ can be significant only 
for much higher redshift. The 95$\%$C.L. level can be well fitted 
by $\Omega_{\rm x} \leq 0.85+(1-4c_s^2/3)(w_0+1)$.  
The constraint on the cosmological constant model($w_0=-1$) 
is less stringent than Kochanek's because 
of the difference in the estimated number density of E/S0 type 
galaxies \cite{ellis}. Taking the same number
density of galaxies as Kochanek's, we obtained the 
constraint $\lambda_0 < 0.67$(95$\%$C.L.)  for flat models 
 consistent with his constraint:  $\lambda_0 < 0.66$
(95$\%$C.L.).

\section{Perturbations in x-matter models}

\subsection{Evolution and General Features}

The gauge invariant perturbation equations for generalized fluid, i.e.,
x-matter are given in the Fourier space as \cite{kodama} 
\begin{equation}
\left({\Delta_{\rm x} \over 1+w_{\rm x}}\right)' = -kV_{\rm x} +
{3 \over 1 + w_{\rm x}}{a' \over a} {1 \over \rho} 
\sum_\alpha c_\alpha^2 \rho_\alpha \Delta_\alpha ~,
\label{pert1}
\end{equation}
and 
\begin{equation}
V_{\rm x}' = -{a' \over a} (1-3c_{s}^2) V_{\rm x}
+c_{s}^2 \left({k \over 1 + w_{\rm x}}\Delta_{\rm x}-
3 {a' \over a}V\right)
+ k \Psi ~,
\label{pert2}
\end{equation}
where $\Delta_{\rm x}$, $V_{\rm x}$ are density and velocity
perturbations of x-matter, $k$ is the wave number, $c_{s}^2$ and $w_{\rm
  x}\equiv p_{\rm x}/\rho_{\rm x}$ are sound speed and pressure-density
ratio of x-matter, $'$ is the derivative in terms of conformal time
$\eta$, $c_\alpha^2$, $\rho_\alpha$ and $\Delta_\alpha$ are sound speed,
density, density perturbations of the $\alpha$ component, respectively,
$\sum_\alpha$ indicates summation in terms of all components, $V$ is
total velocity perturbations and $\Psi$ is the gravitational potential.
Above density perturbations are ones relative to the total matter rest
frame.  We solve these equations numerically together with equations for
ordinary baryonic matter, radiation (photons and neutrinos) 
and/or cold dark matter.  The time
evolutions of x-matter density fluctuations are shown in Figure 5.  In
this figure, we take $h=0.5, \Omega_B=0.05, \Omega_{\rm x}=0.95$ and
$\Omega_{\rm CDM}=0$.  For the comparison, we plotted the evolution of
CDM perturbations (a dotted line).

If we set $c_s^2$ to be zero in order to study the dependence on $w_0$,
we find decay of fluctuations near the present epoch (a dashed line).
It is interesting to examine what causes this turnover.  In case of
CDM dominated models, we can assume $\Psi$ is constant in time, and the
scale factor $a \propto \eta^2$. Therefore from equations (\ref{pert1}) 
and (\ref{pert2}),
we obtain $V_{\rm x} \propto \eta$ and $\Delta_{\rm x} \propto \eta^2
(1+w_{\rm x})$.  Accordingly, we find $\Delta_{\rm x}' \propto \eta (1 +
w_{\rm x})(3 w_{\rm x}+1)$.  Thus, we can conclude that 
$\Delta_{\rm x}$ decays if $w_{\rm x} < -1/3$ in case of CDM  
dominated models.  This critical value $w_{\rm x} = -1/3$ corresponds to
the breaking of the strong energy condition.  
For x-matter dominated models, we numerically find a 
slightly larger critical value ($w_{\rm x} \simeq -0.28$).
It is because even for higher $w_{\rm x}$ in these models, 
decay of the self-gravitational potential makes perturbations
difficult to grow.

When we take a non-zero value of $c_s^2$, on the other hand, 
we find the acoustic oscillation of fluctuations of x-matter 
after crossing the Jeans scale.
In Figure 6, the Jeans mass scales for various $\Omega_{\rm x}$ and
$c_s^2$ are shown as a function of redshift.  The Jeans mass of
x-matter is defined as
\begin{equation}
M_{\rm J} = {4 \pi \over 3} \rho_{\rm x}
\left({\lambda_{\rm J} \over 2}\right)^3 ~ ,
\end{equation}
where 
\begin{equation}
\lambda_{\rm J} = \left({\pi c_{s} \over G\rho_{\rm tot }}\right)^{1/2} ~. 
\end{equation}
Here $\rho_{\rm tot}$ is the total density defined as 
$\rho_{\rm tot} = \rho_{\rm x}+\rho_{\rm CDM} + \rho_{\rm B}$.
Solid and dashed lines denote $w_0=-0.2$ and $-0.8$, respectively. 
In this figure, we find that it is difficult to form the structure 
below galaxy scales if $c_s^2 > 10^{-8}$ when x-matter is the dominant
component of the universe. However, the constraint can be loosened if we
allow the time variation of $c_s^2$.

To see the effect of $w_0$ and  Jeans oscillation, the matter transfer
functions are shown in Figure 7.  In panel (a), we take $c_s^2=0$.  In
this panel, smaller $w_0$ indicates smaller amount of x-matter in past
and later matter-radiation equality since we fix the present
x-matter density.  Therefore the location of the knee in the transfer
function which corresponds to the horizon scale of the matter-radiation 
equality shifts to the larger scale when we take smaller $w_0$. 
In panel (b), on the other hand, we fix $w_0$ to be $0$ and change $c_s^2$.
The Jeans oscillation is seen in the transfer function even in case of 
$c_s^2=10^{-5}$.  

In Figure 8, the angular power spectra of CMB anisotropies, so-called
$C_\ell$ are shown for the same models as Figure 5.  The shape of
$C_\ell$ for the $c_s^2=0$ model is quite similar to the one with the
cosmological constant.  The peak locations and the shape on small $\ell$
(large scales) of this model are different from sCDM which is the
fiducial CDM models with $h=0.5, \Omega_0=1$ and $\Omega_B=0.05$.  A 
dent on the large scale ($\ell\sim 15$) is caused by the decay of
gravitational potential just like lambda models (see e.g., 
Sugiyama \& Silk 1994).  
Since the volume of
the universe of the negative $w_0$ model is bigger than the one of the
standard Einstein-de Sitter model, the peak locations shift to the
smaller scales as is shown in this figure.

On the other hand, $C_\ell$ of the model with high $c_s^2$ 
has a quite different shape.  In this model, the universe has never 
become purely matter dominated or radiation dominated.  
In other words, the gravitational potential has been decaying all the 
time.  This decay boosts $C_\ell$ in the long range of $\ell$'s.

One interesting feature of this figure is that the damping locations at
high $\ell$ of all $C_\ell$'s are almost the same.  It is simply because
recombination history is the same regardless of the equation of state
since it is determined by the ratio between the number densities of 
photons and baryons.

\subsection{Observational Constraints}

Now let us make a comparison with observational data.  In Figures 9
($h=0.5$) and 10 ($h=0.7$), we plot the contour maps of $\sigma_8$, which
is the rms density field in a sphere of radius $8h^{-1}$Mpc with the top
hat window function, on the $w_0-\Omega_{\rm x}$ plane.  Here we employ
COBE 4yr normalization~\cite{Benn,BW}.  
The observational value of $\sigma_8$ is 
$\sigma_8=(0.52 \pm 0.04)\Omega_{0}^{0.52+0.13\Omega_{0}}$ 
for flat models from the analysis of the local cluster X-ray
temperature function \cite{eke}. 
Here $\Omega_0$ is the density parameter of total matter in case of 
flat $\Lambda$-models.  
Similar results are obtained from other analyses \cite{wef,vl1}.
While there is $\Omega_{0}$
dependence in the observational value of $\sigma_8$, we expect the
dependence of $\sigma_8$ on $\Omega_{\rm x}$ and $w_0$ is relatively
weak unless  $w_0 \simeq -1$ and $\Omega_{\rm x} \simeq 1$.

Employing $\sigma_8 \simeq 0.6$, we plotted total matter transfer
functions in Figures 11 (panels (a)-(c) for $h=0.5$ and panels (d)-(f)
for $h=0.7$) for various $c_s$'s and $w_0$'s.  In case of $c_s^2 <
10^{-5}$, transfer functions do not depend on $c_s^2$ since the Jeans
scale is smaller than $1\rm h^{-1}Mpc$ even at present.  Therefore the
shape of transfer functions is merely controlled by the matter-radiation
equality epoch since the horizon scale at this epoch corresponds to the
knee of the transfer function.  Regarding the normalization of
perturbations, CMB anisotropies are affected by the decay of
gravitational potential, which is so-called integrated Sachs-Wolfe effect.
Hence one might expect
different normalization factors from the same shape of the transfer
function for models with different $w_0$'s.  However, the decay of the
gravitational potential caused by $w_0$ takes place at very late epoch
even in case $w_0 \simeq -1$ as is shown in Figure 5.  Therefore the
integrated Sachs-Wolfe effect does not provide dominant contribution on
CMB anisotropies on the COBE scale unless the universe is dominated by
x-matter.  Eventually we conclude that fixing $\sigma_8$ indicates
fixing the matter-radiation equality and fixing the shape of the
transfer function.  On the other hand, we find a mild damping feature in
the transfer function on small scales just like a mixed dark matter
model(i.e. cold+hot dark matter) if we take $c_s^2 \geq 10^{-4}$ 
because of the Jeans oscillation of
x-matter perturbations.  In particular, it is interesting that models
with $\Omega_{\rm x}\simeq 0.1$, $c_s^2=0.15$ and $w_0=-0.8$ fit
fairly well to the observational result by Peacock and
Dodds \shortcite{pd} as shown in Figure 12.
 
In Figure 13 ($h=0.5$ for panels (a)-(c) and $h=0.7$ for panels
(d)-(f)), we plot $C_\ell$'s of models with $\sigma_8=0.6$.  In case of
$c_s^2 < 10^{-5}$, $C_\ell$'s do not depend on neither $c_s$, $w_0$ nor
$\Omega_{\rm x}$ (see panel (a) or (d)).  Comparing to the pure CDM
model, however, we find higher acoustic peaks.  It is because the epoch
of the matter-radiation equality is the same for models with the same
$\sigma_8$.  On the other hand, the equality epochs of the pure CDM models
whose $\sigma_8$ are $1.2$ and $1.6$ for $h=0.5$ and $0.7$,
respectively, are earlier than ones of x-matter models.  
It is well known that the later the equality
epoch, the higher the peaks are \cite{hss}.
It is the same for panels (b) or (e).  Smaller $w_0$
indicates smaller amount of x-matter in past and the later matter
radiation equality since we fix the present x-matter density.  Therefore
it is shown higher peaks for smaller $w_0$.  In case of larger $c_s^2$,
on the other hand, 
the peak locations are shifted on smaller scales (larger $\ell$) as is
shown in panels (c) or (f).  With larger $c_s^2$, the
equation-of-state of x-matter is  
in between radiation and matter and x-matter is a (sub)-dominant component 
in the universe around the matter-radiation equality epoch even for 
small  $\Omega_{\rm x}$.
This component causes the decay of the gravitational potential since
only pure radiation or matter domination keeps the potential staying
constant. And this decay boosts the first acoustic peak as is shown in 
panels (c) or (f). 

In conclusion, adopting $\sigma_8\simeq 0.6$ constrains x-matter
content $\Omega_{\rm x} \siml 0.1$ for $c_s^2 \simeq 0.1$, or if we
consider the universe to be x-matter dominated, then its sound speed
is constrained as  $c_s^2 < 10^{-5}$ at least at the epoch of galaxy
formation. Former case can fit the observational data by Peacock-Dodds 
well. With  high precision measurements expected from
MAP\footnote{http://map.gsfc.nana.gov}(Microwave
Anisotropy Probe), PLANCK\footnote{http:
//astro.estec.esa.nl/SA-general/Projects/Planck}, 
2DF\footnote{http://meteor.anu.edu.au/$\sim$colless/2dF}(2 Degree
Field) and  
SDSS\footnote{http://www.astro.prinveton.edu/BBOOK}(Sloan
Digital Sky Survey), we should distinguish x-matter models from 
sCDM or $\Lambda$CDM.

\section{summary}

We have investigated the possibility that the dark matter component has
the equation of state $p_x=w_0\rho_{x0}+c_s^2(\rho_x-\rho_{x0})$ such
that $-1\leq w_0 < -1/3$ and $0 \leq c_s^2 \siml 0.15$.  We have
studied the detailed observational consequences of x-matter models 
including the
statistics of gravitational lensing of quasars, matter power spectrum
and cosmic microwave background (CMB) anisotropies together with 
a comprehensive study of density perturbations of x-matter models. 
The gravitational lensing statistics constrains $\Omega_{\rm x}$ as 
$\Omega_{\rm x} \leq 0.85+(1-4c_s^2/3)(1+w_0)$(95$\%$C.L.). 
We have shown that introducing the sound speed of x-component has a
great effect on CMB anisotropies and matter power spectrum.  
We find that adopting $\sigma_8\simeq 0.6$ constrains x-matter
content $\Omega_{\rm x} \siml 0.1$ for $c_s^2 \simeq 0.1$, which can 
fit the observational data by Peacock-Dodds well. Alternatively, if we
consider the universe to be x-matter dominated, then its sound speed
is found to be constrained as  $c_s^2 < 10^{-5}$ at least at the epoch 
of galaxy formation. Although x-matter models are consistent with all 
current observations, they leave imprints in the CMB anisotropy and 
matter power spectrum that should be detectable in future
observations.

\section*{Acknowledgments}

We would like to thank Dr. D.Maoz for providing us with the HST
Snapshot Survey data. 
This work was supported in part by a 
Grant-in-Aid for Basic Research of the Ministry of Education,
Culture, and Sports Nos. 09440106(NS) and 09NP0801(TN). 
One of the authors (TC) is supported by JSPS Research Fellowships for 
Young Scientists.

\appendix

\section{effective potential for x-matter model}
\label{sc:poten}

We can construct the effective potential of a scalar field for the
x-matter model. The method of the construction is similar to that
given in Nakamura \& Chiba\shortcite{nc}.

We consider a scalar field $\phi$ minimally coupled to gravity. Then the
energy density and the pressure of x-matter are given in terms of the
scalar field as
\beqa
\rho_{\rm x}&=&{1\over 2}\dot{\phi}^2-V(\phi),\\
p_{\rm x}&=& {1\over 2}\dot{\phi}^2+V(\phi),
\eeqa
where $V(\phi)$ is the potential of the scalar field. 
By the use of Eq.(\ref{rho}) and Eq.(\ref{eos}), the scalar field and
the potential can be written in terms of the scale factor
\beqa
2V(a)&=& \rho_{\rm x}-p_{\rm x}={2(c_s^2-w_0)\over 1+c_s^2}\rho_{\rm x0} +
{(1-c_s^2)(1+w_0)\over 1+c_s^2}\rho_{\rm x0}a^{-3(1+c_s^2)},\\
\dot{\phi}^2&=&\left(aH {d\phi\over
da}\right)^2=(1+w_0)\rho_{\rm x0}a^{-3(1+c_s^2)}.
\eeqa
$\phi$ is given by the integral 
\beq
\phi(a)-\phi_0=\pm \int ^a_1\sqrt{3(1+c_s^2)(1+w_0)\over 8\pi G 
\left( 1+w_0 +(c_s^2-w_0)u^{3(1+c_s^2)}\right)
}{du\over u}.
\eeq
Then $V$ is given implicitly as a function of $\phi$. The shapes of 
$V(\phi)$ for $c_s^2=0.0,0.1$ are shown in Fig.14 from $a=1$ to
$a=10^{-3}$.  The scale is arbitrary. Since $V(\phi)$ involves a tiny
(but positive finite) constant term, the universe will be dominated by 
this constant term and behave like the cosmological constant dominated 
universe in future.

\section{formulae for a singular isothermal sphere lens}
\label{sc:tog}

Here we give useful formulae for a singular isothermal sphere lens 
which are necessary in the text. 

The deflection angle $\alpha$ by  an  isothermal sphere lens 
is given by
\beq
\alpha={4GM(\leq b)\over b}.
\eeq
Here $b$ is the impact parameter and $M(\leq b)$ is the mass
projected within the impact parameter $b$
\beq
M(\leq b)=2\pi \int_0^b \Sigma(r) rdr.
\eeq
$\Sigma(r)$ is the surface density and is written for a singular
isothermal sphere 
in terms of the
velocity dispersion $v$ as
\beq
\Sigma(r)={v^2\over 2Gr}.
\eeq
Therefore the deflection angle is written as
\beq
\alpha=4\pi v^2.
\eeq
The lens equation is then
\beq
\phi=\theta-4\pi v^2{D_{LS}\over D_{OS}},
\eeq
where $\phi$ and $\theta$ are the angular position of the source and
the  image, respectively. The critical impact parameter is thus
\beq
b_{cr}=4\pi v^2{D_{OL}D_{LS}\over D_{OS}}.
\label{crit}
\eeq
Therefore, the cross section is given by 
\beq
\sigma=\pi b_{cr}^2=16\pi^3v^4\left({D_{OL}D_{LS}\over
D_{OS}}\right)^2.
\eeq

If the alignment is close enough such that $\phi \leq \alpha{D_{LS}\over 
D_{OS}}$,  lensing produces multiple images at 
\beq
\theta^{\pm}=\phi \pm \alpha{D_{LS}\over D_{OS}}, 
\eeq
An amplification due to lensing is given by
\beq
A=\Bigl{|}{\theta d\theta\over \phi d\phi}\Bigr{|}.
\eeq
The two images are amplified by factors
\beq
A^{\pm}={\alpha D_{LS}\over \phi D_{OS}}  \pm 1.
\eeq
This gives a total amplification
\beq
A=2{\alpha D_{LS}\over \phi D_{OS}}. 
\eeq
This means that the probability distribution of total amplification
for multiply imaged systems is
\beq
P(A)dA=8A^{-3}dA,~~~~~~~~~{\rm for}\ A\geq 2.
\eeq

\bsp

\section*{Figure Captions}

\vspace*{12pt}
\noindent
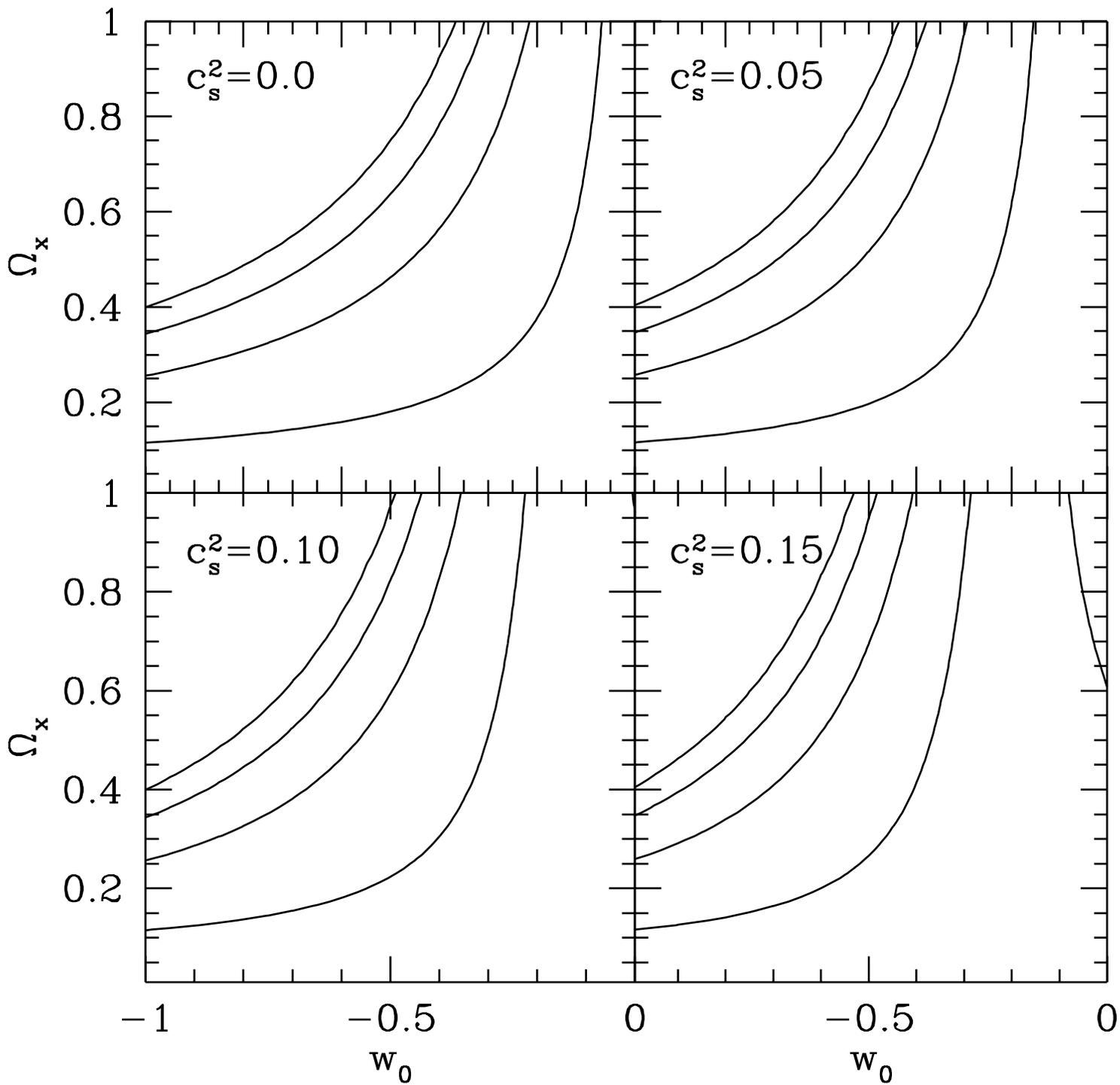
{\bf Figure 1: }
The contour plots for $H_0t_0$ in the $(w_0,\Omega_{\rm x})$ plane. 
The contour levels are $H_0t_0=1.0,0.9,0.8,0.7,0.6$(only for 
$c_s^2=0.15)$ from top-left to right.

\vspace*{12pt}
\noindent
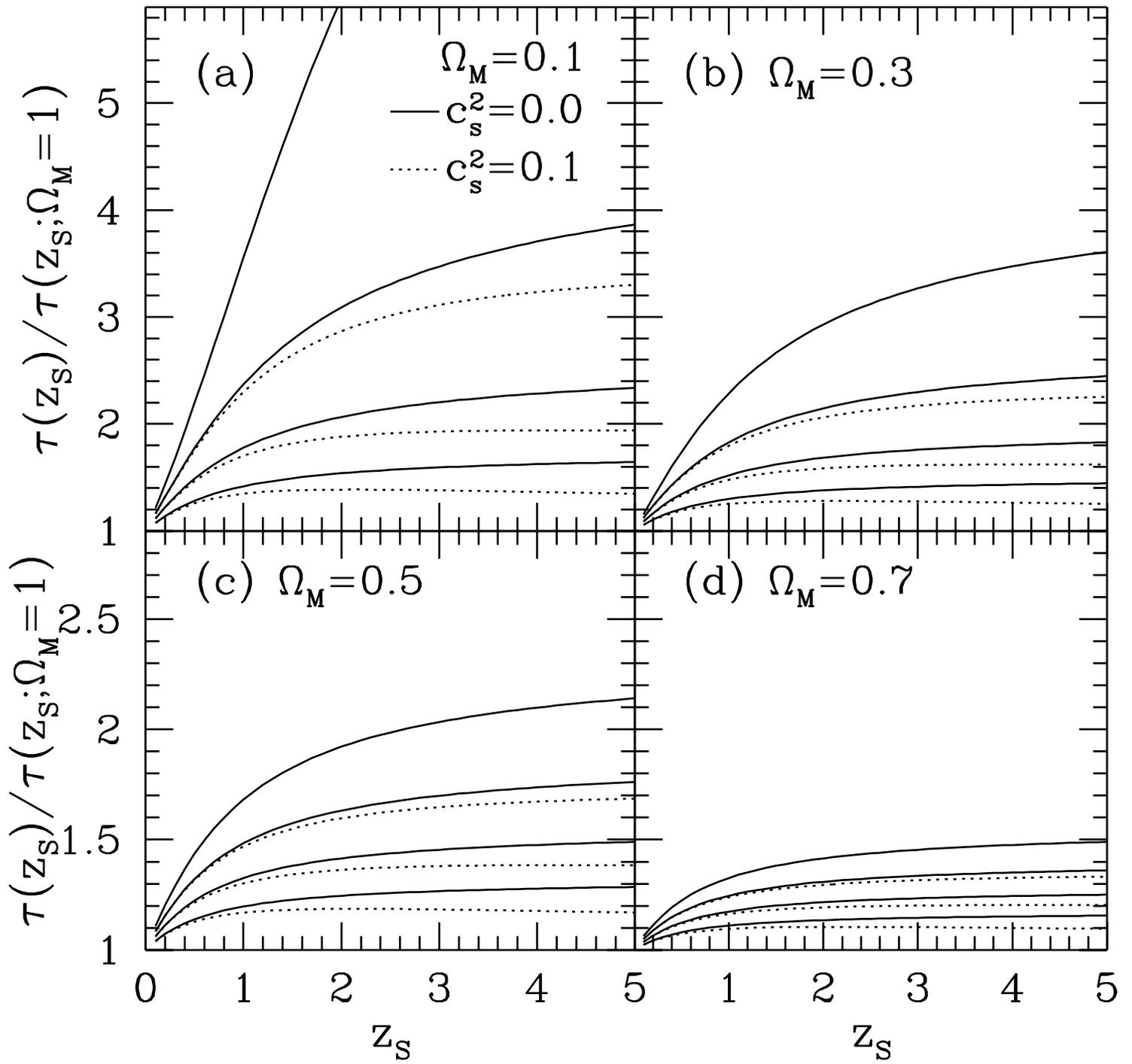
{\bf Figure 2: }
The lens optical depth for a source at redshift $z_s$ in a flat
universe relative to 
that in the Einstein-de Sitter model($\Omega_{M}=1$)
for (a)$\Omega_{M}=0.1$, (b) $\Omega_{M}=0.3$, (c) $\Omega_{M}=0.5$, and
(d) $\Omega_{M}=0.7$. Solid curves correspond
to $c_s^2=0$, dotted ones $c_s^2=0.1$.
Each type of curves from top to bottom corresponds 
to $w_0=-1,-0.8,-0.6,-0.4$. Note that $w_0=-1$ corresponds to the 
cosmological constant model irrespective of $c_s^2$. 

\vspace*{12pt}
\noindent
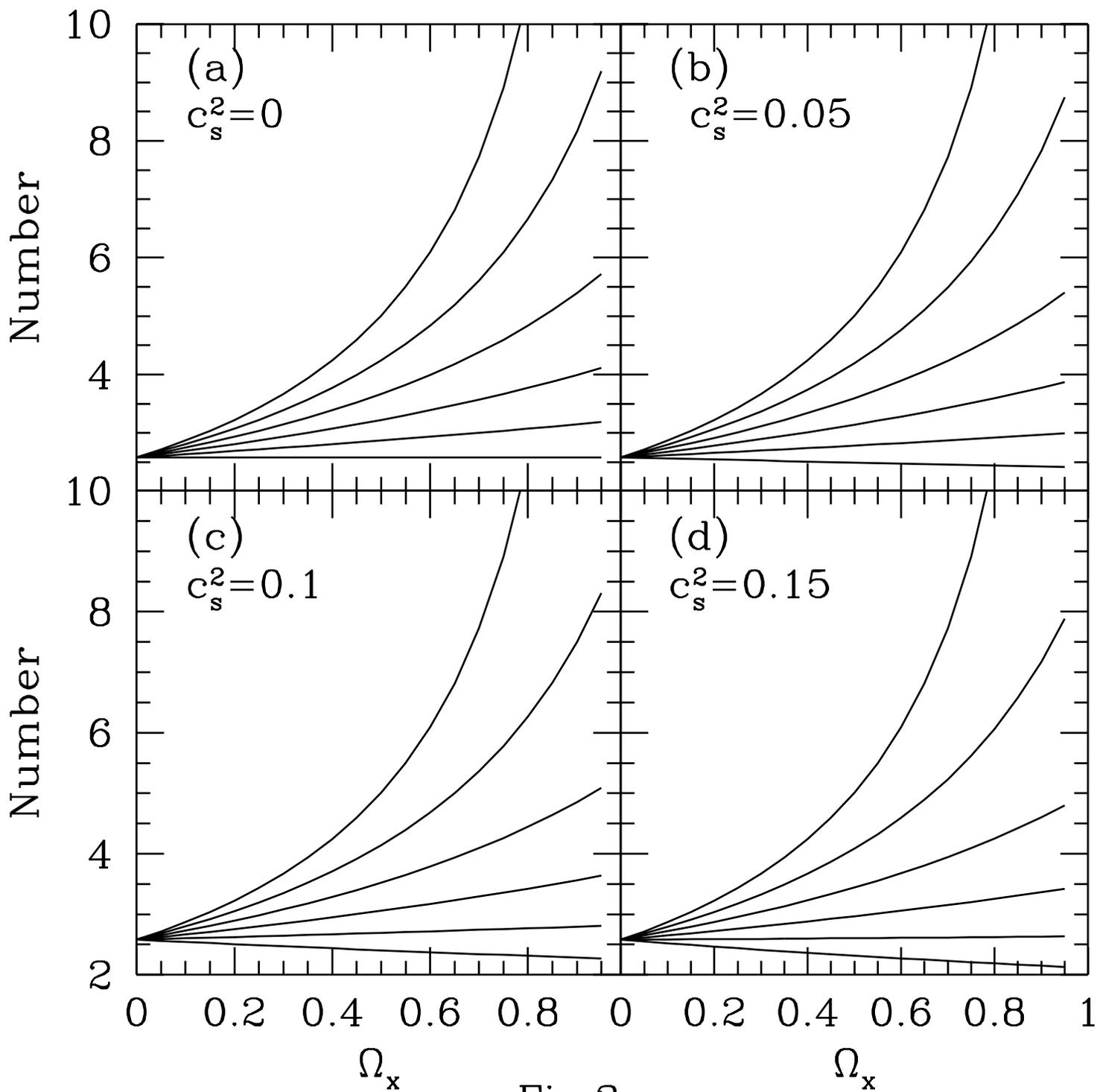
{\bf Figure 3: }
The expected number of lenses for $c_s^2=0$(a),0.05(b),0.1(c),0.15(d) 
and $w_0=-1,-0.8,-0.6,-0.4,-0.2,0$ as a function of $\Omega_{\rm x}$. 
Each  curve from top to bottom corresponds 
to $w_0=-1,-0.8,-0.6,-0.4,-0.2,0$. $w_0=-1$ corresponds 
to the cosmological constant model irrespective of $c_s^2$. 
Data is taken from HST snapshot survey\cite{maoz}. 

\vspace*{12pt}
\noindent
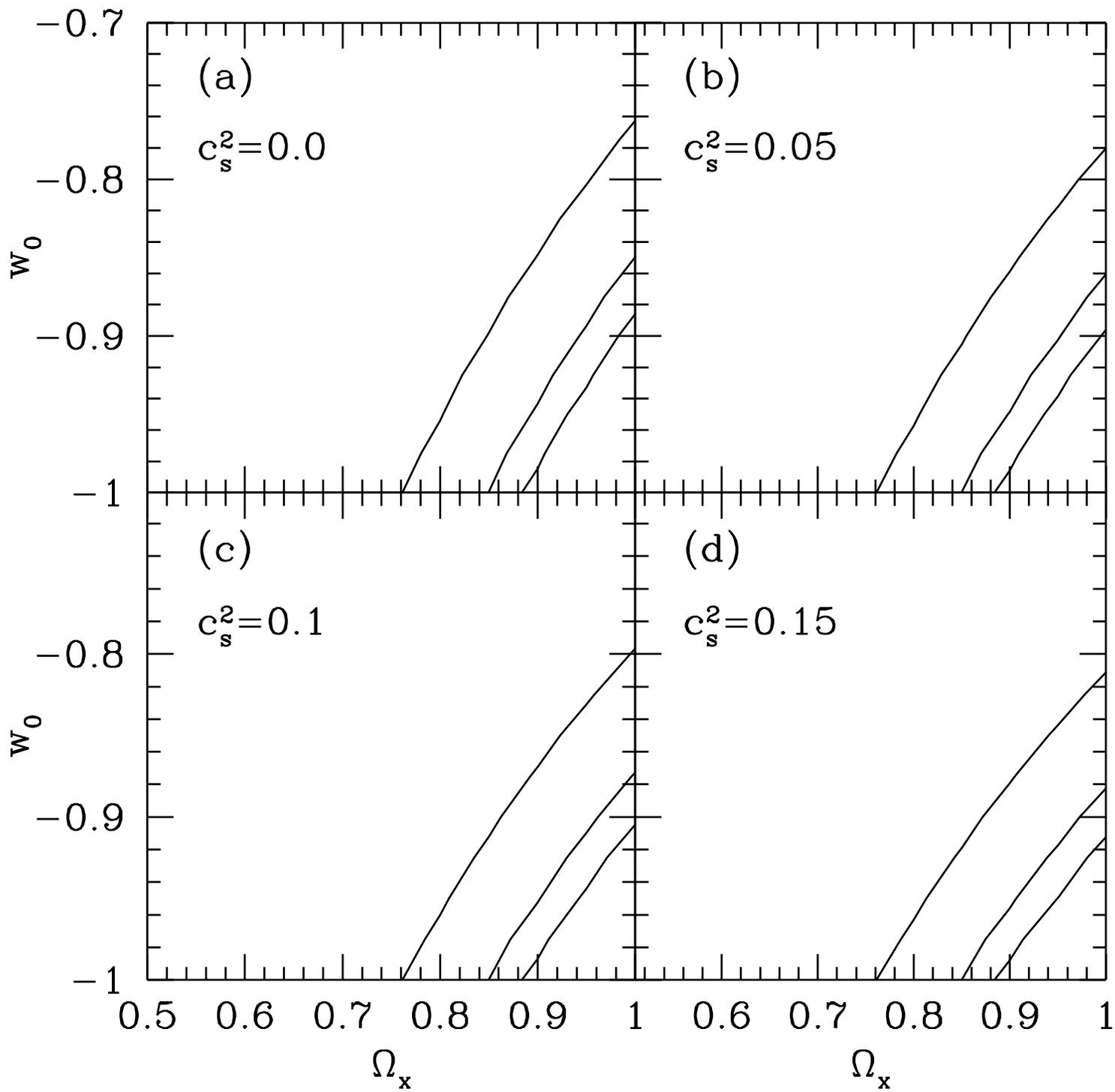
{\bf Figure 4: }
Likelihood contours for  flat cosmological models projected 
onto constant $c_s$ plane. 
Likelihood levels are, from left top
to right down, $68.3\%, 95\%, 99\%$.

\vspace*{12pt}
\noindent
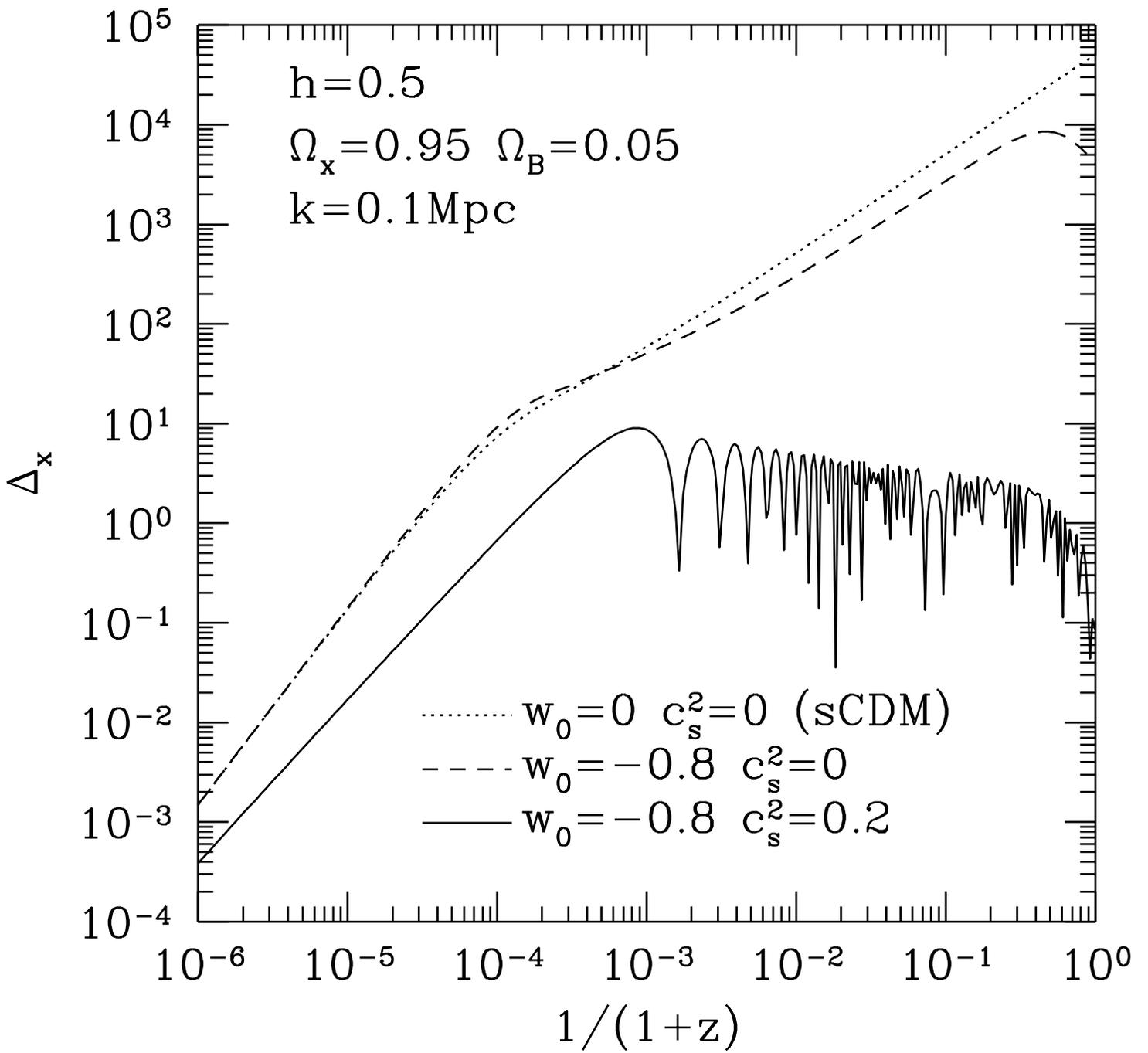
{\bf Figure 5: }
Time evolution of x-matter density fluctuations $\Delta_{\rm x}$
at the wave number $k=0.1$Mpc for different values of $w_0$ and
$c_s^2$.  Solid and dotted lines are for $w_0=-0.8$, $c_s^2=0.2$ and
$w_0=-0.8$, $c_s^2=0$, respectively.  A dashed line corresponds to the
standard CDM model, i.e., $w_0=0$ and $c_s^2=0$.  We take 
$\Omega_{\rm x}=0.95$ and $\Omega_B=0.05$ as cosmological model parameters.  

\vspace*{12pt}
\noindent
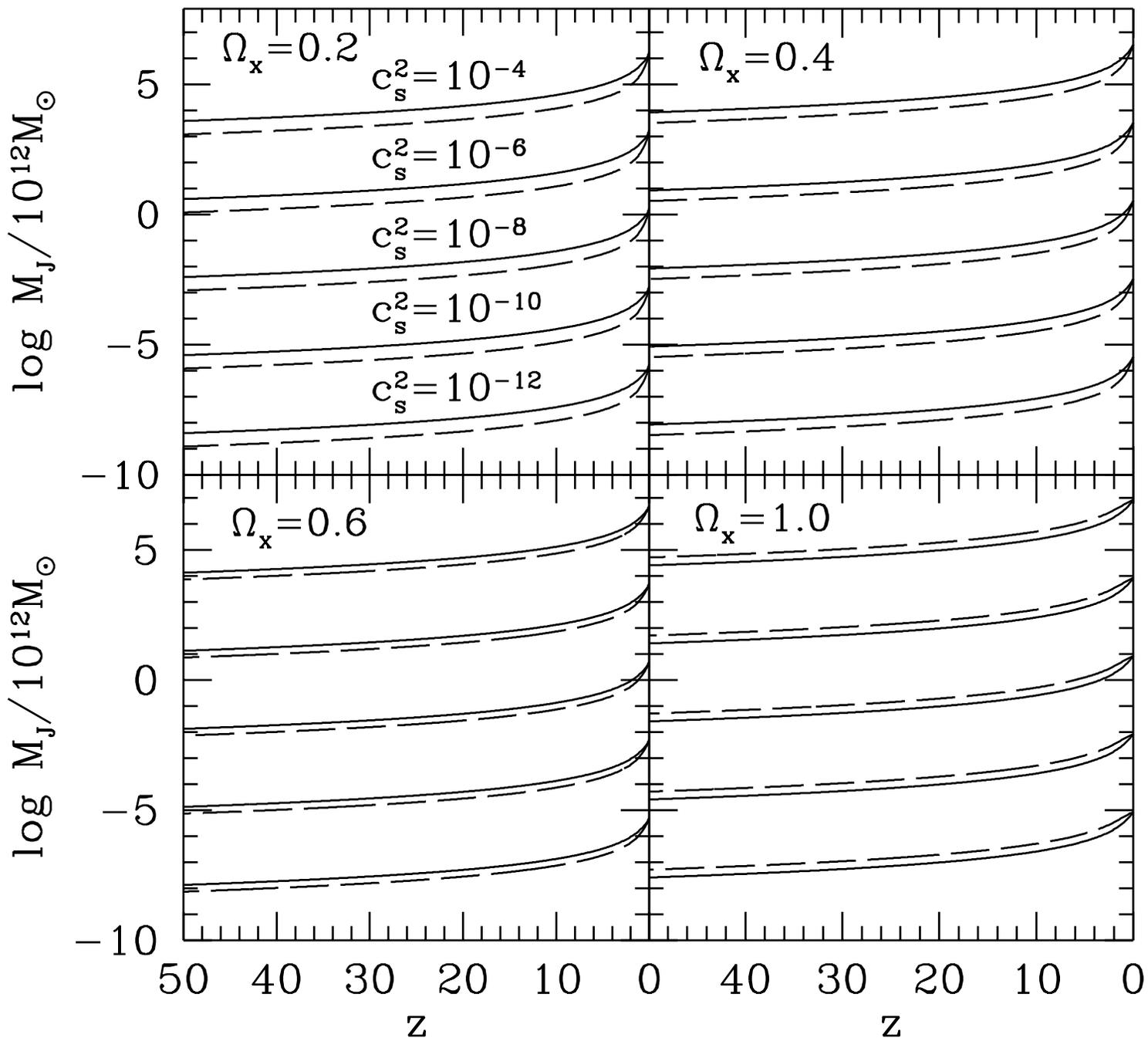
{\bf Figure 6: }
Jeans scale as a function of redshift. 
Solid lines correspond to $w_0=-0.2$, while
dashed lines to $w_0=-0.8$.

\vspace*{12pt}
\noindent
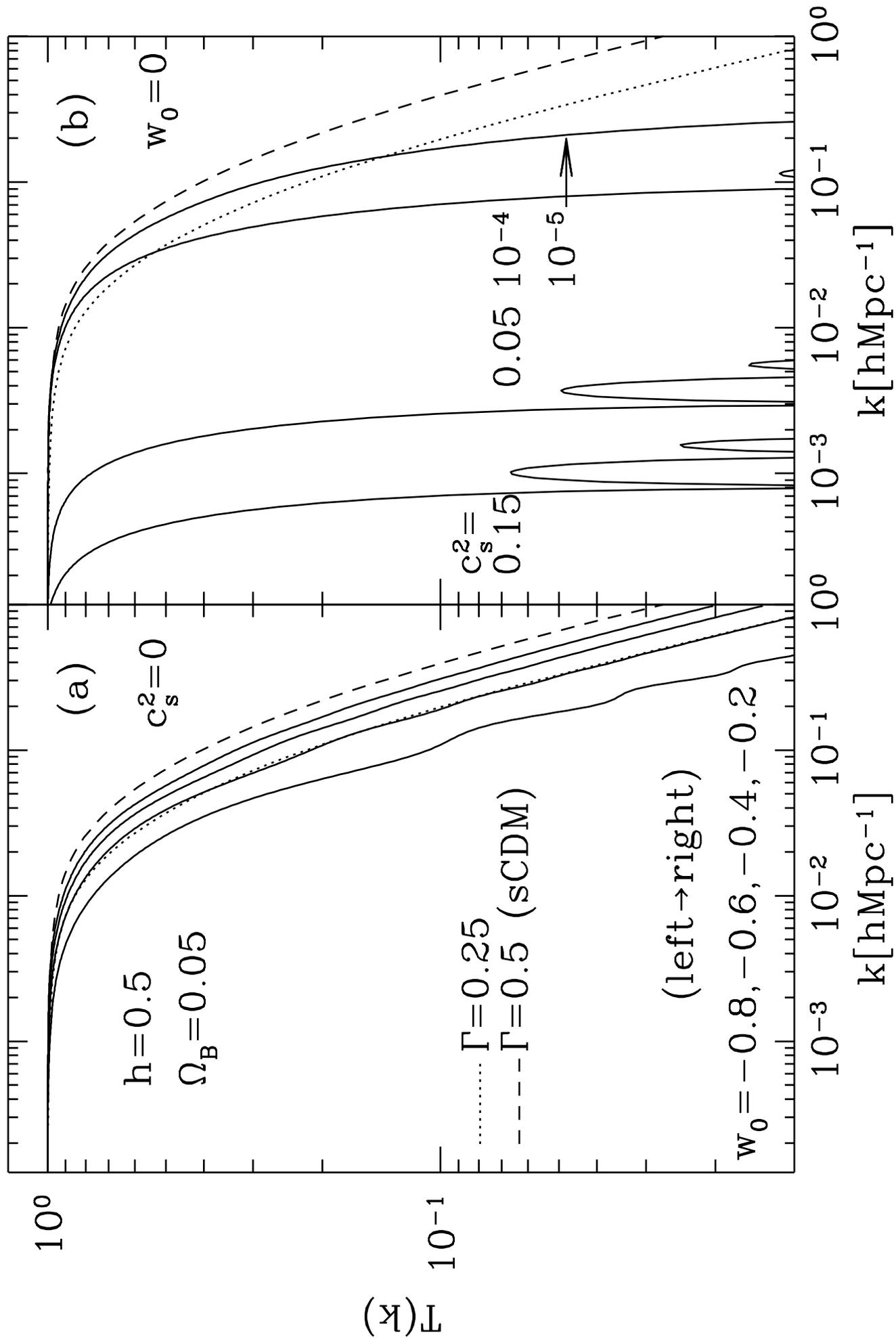
{\bf Figure 7: }
Total matter transfer functions.  We take 
$\Omega_{\rm x}=0.95$ and $\Omega_B=0.05$ as cosmological model parameters.
(a): changing $w_0$ with $c_s^2=0$.  From left to right, $w_0=-0.8,
-0.6, -0.4$ and $-0.2$.  Dashed and dotted lines are CDM models with the
shape parameter $\Gamma=0.5$ and $0.25$, respectively.  (b): changing
$c_s^2$ with $w_0=0$.  From left to right, $c_s^2=0.15, 0.05, 10^{-4}$
and $10^{-5}$.

\vspace*{12pt}
\noindent
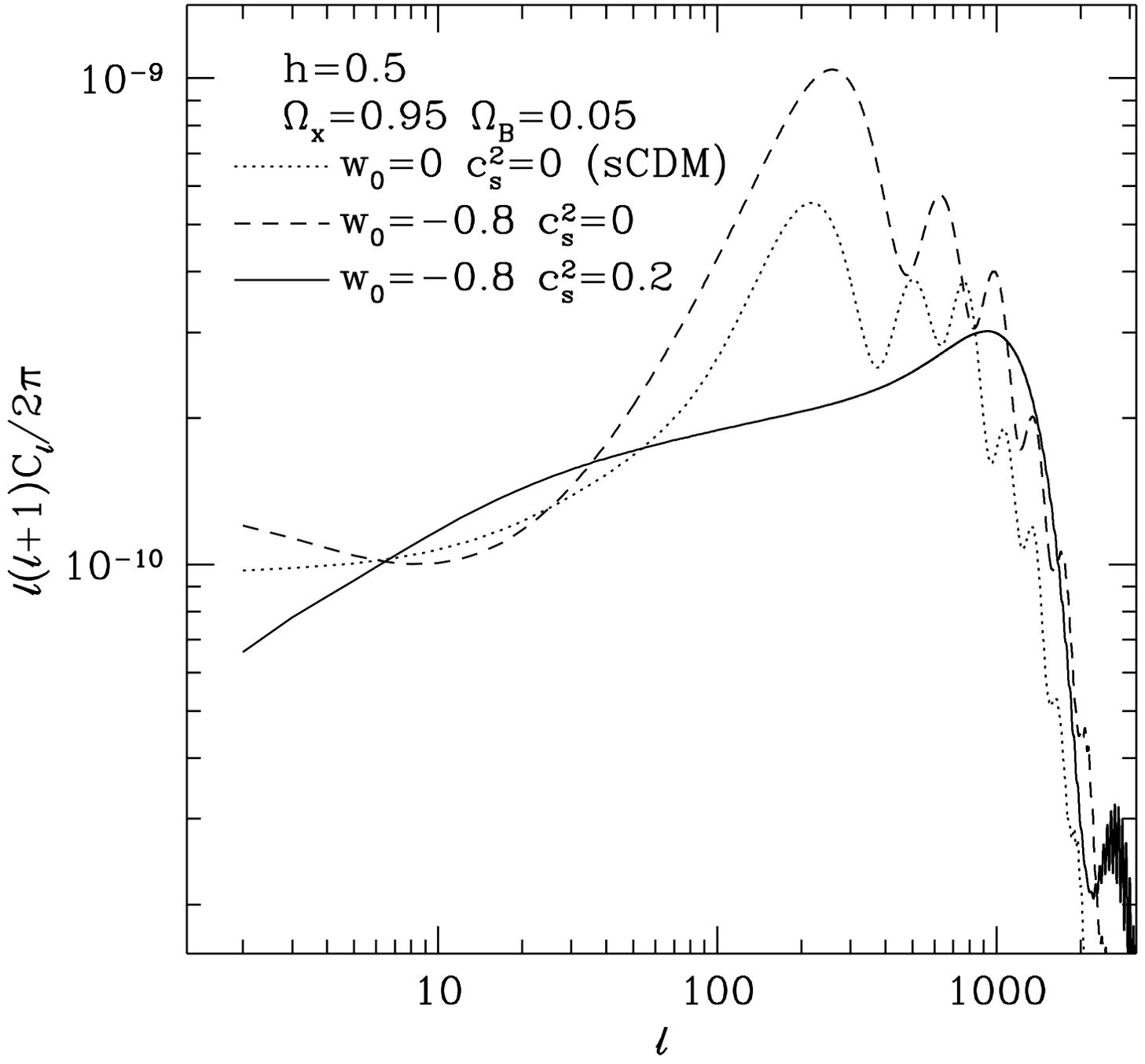
{\bf Figure 8: }
CMB angular power spectra.  Models are same as Figure 5. 

\vspace*{12pt}
\noindent
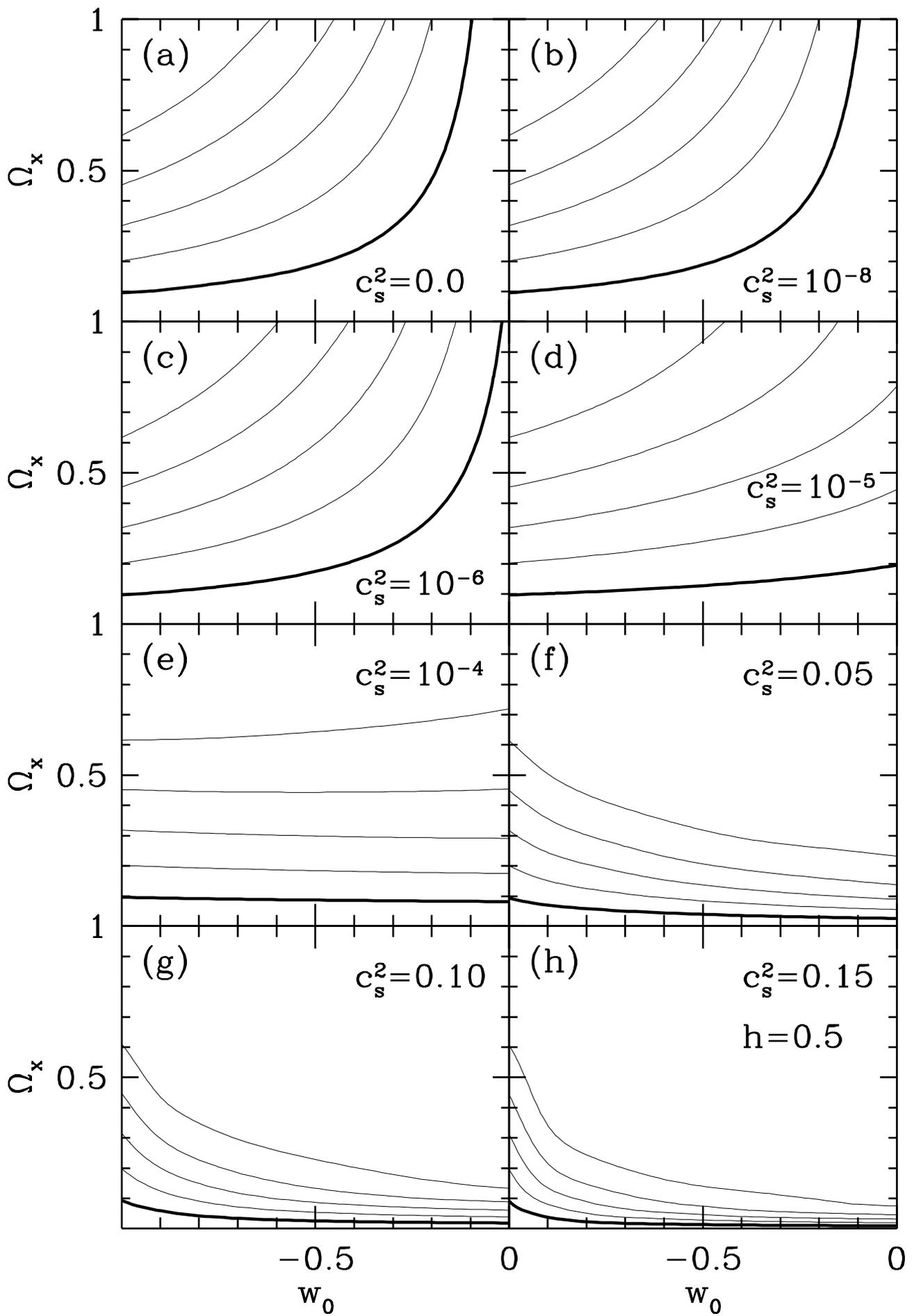
{\bf Figure 9: }
Contours of $\sigma_8$ in the $w_0-\Omega_{\rm x}$ plane for
various values of $c_s$.  COBE 4yr normalization is employed.
We take  $h=0.5$ and $\Omega_B=0.05$.  From top to bottom
in each panels, $\sigma_8=0.2, 0.4, 0.6, 0.8$ and $1.0$ (bold).

\vspace*{12pt}
\noindent
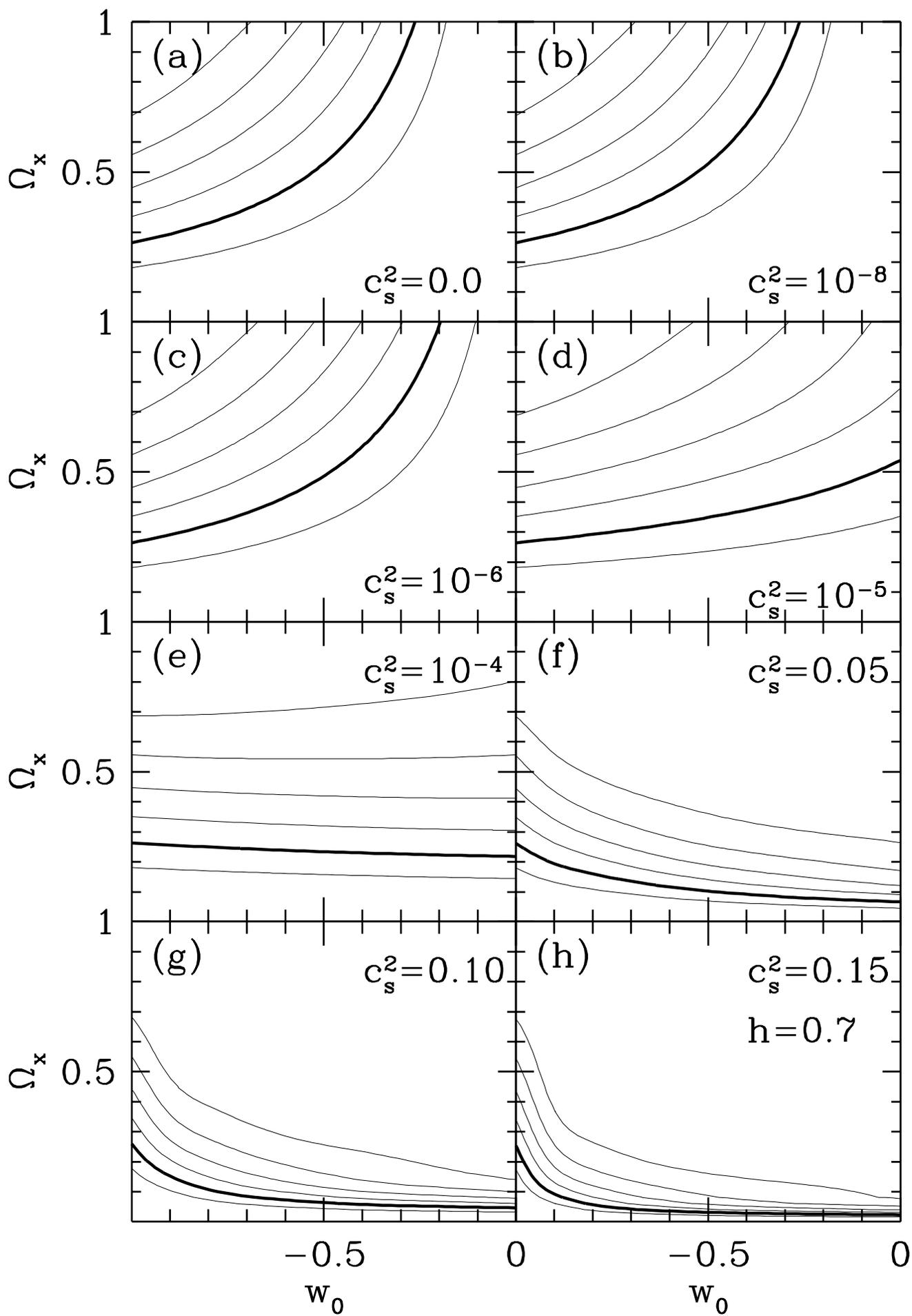
{\bf Figure 10: }
Same as Figure 9 but $h=0.7$ and $\Omega_B=0.04$.

\vspace*{12pt}
\noindent
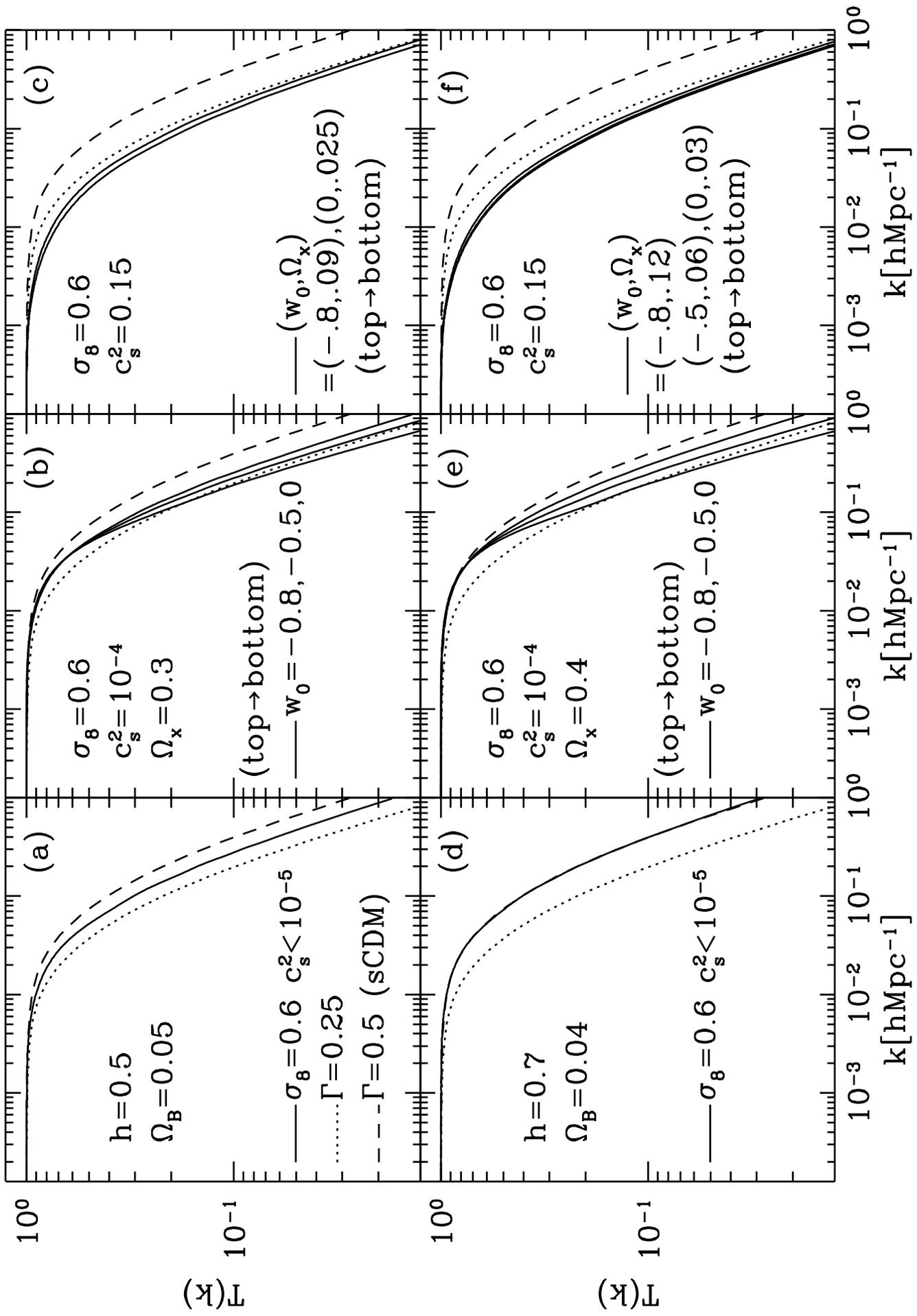
{\bf Figure 11: }
Total matter transfer functions for models with $\sigma_8=0.6$.
Panels (a)-(c) are models with $h=0.5$ and $\Omega_B=0.05$ and Panels
(d)-(f) are with $h=0.7$ and $\Omega_B=0.05$.  Solid lines are indicated
in each panels.  Dashed and dotted lines are CDM models with the shape
parameter $\Gamma=0.5$ and $0.25$, respectively.  Since the transfer
functions of the model with $w_0=0$ and $\Omega_{\rm x}=0.025$ and the one
with $w_0=-0.5$ and $\Omega_{\rm x}=0.04$ are identical in panel (c), we do
not plot the later one.

\vspace*{12pt}
\noindent
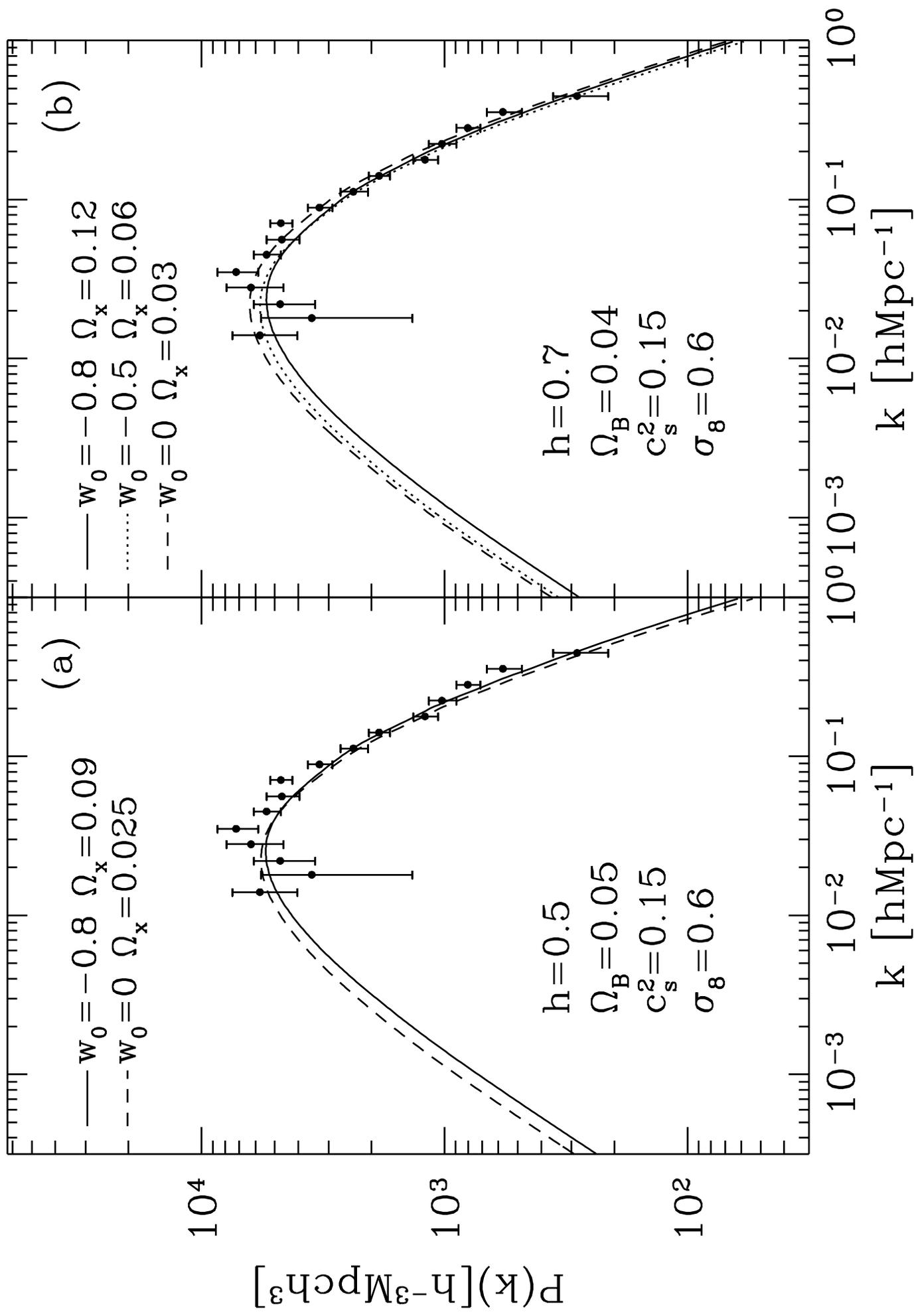
{\bf Figure 12: }
Matter power spectra for `best fitted' models.  From Figure 11,
we chose models with transfer functions similar to the ones of
$\Gamma=0.25$.  (a): models with $h=0.5$ and $\Omega_B=0.05$. Solid and
dashed lines are models with $w_0=-0.8$ and $\Omega_{\rm x}=0.09$ and $w_0=0$
and $\Omega_{\rm x}=0.025$.  (b): models with $h=0.7$ and $\Omega_B=0.04$.
Solid, dotted and dashed lines are models with $w_0=-0.8$ and
$\Omega_{\rm x}=0.12$, $w_0=-0.5$, $\Omega_{\rm x}=0.06$ and $w_0=0$ and
$\Omega_{\rm x}=0.03$, respectively.  The symbols denote the observational data 
by Peacock and Dodds\shortcite{pd}.  We multiply $0.8$ to these
observational values since we are only interested in the shape of the
matter power spectrum and Peacock and Dodds' data provide roughly 
$10-20\%$ larger $\sigma_8$ compared to the value quoted in the text.

\vspace*{12pt}
\noindent
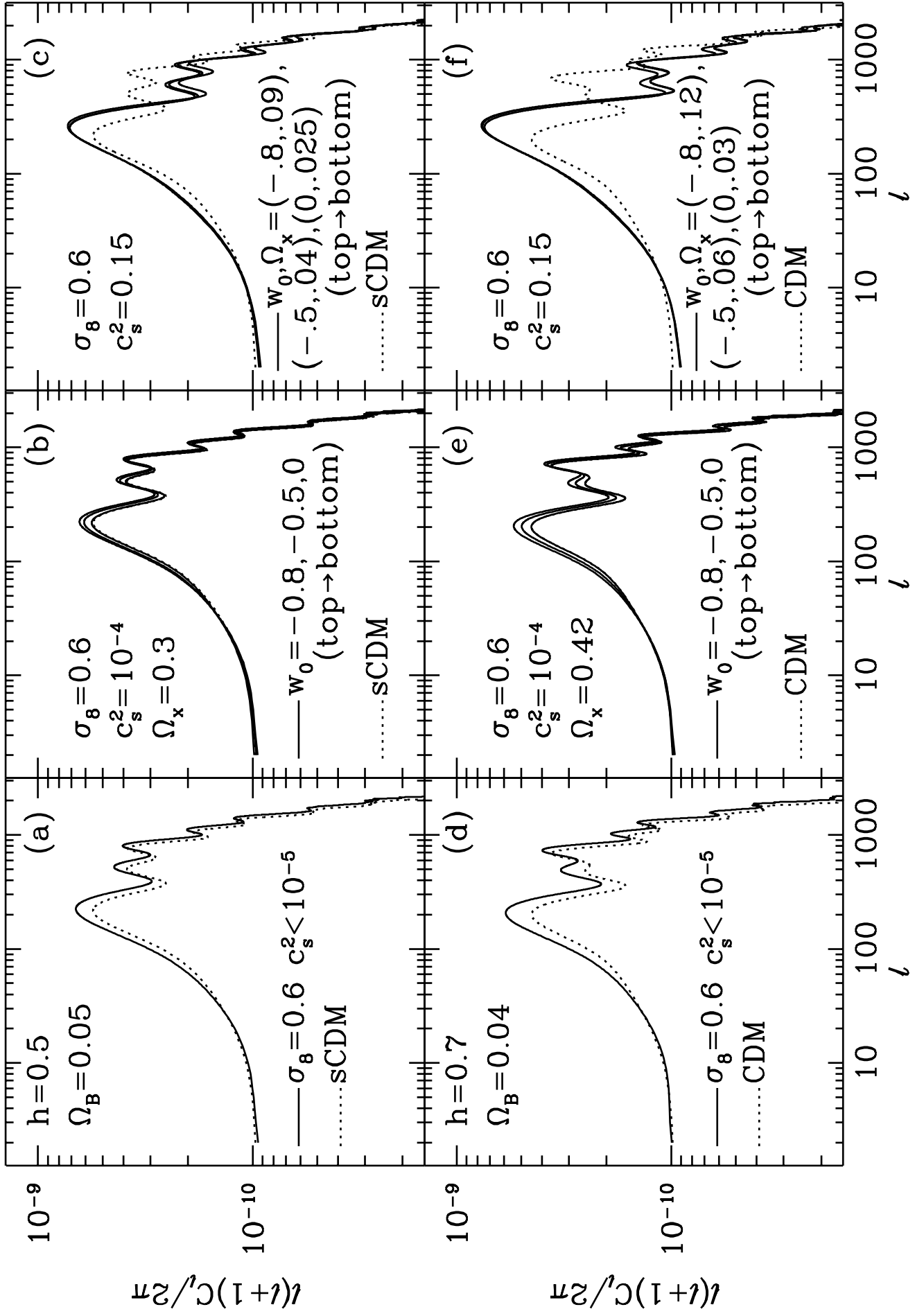
{\bf Figure 13: }
CMB angular power spectra for models with $\sigma_8=0.6$.
Models are same as Figure 11. Dotted lines are $C_\ell$'s of CDM models.
Namely CDM models with $\Omega_0=1, h=0.5$ and  $\Omega_B=0.05$ for 
panels (a)-(c), and 
with $\Omega_0=1, h=0.7$ and  $\Omega_B=0.04$ for 
panels (d)-(f).

\vspace*{12pt}
\noindent
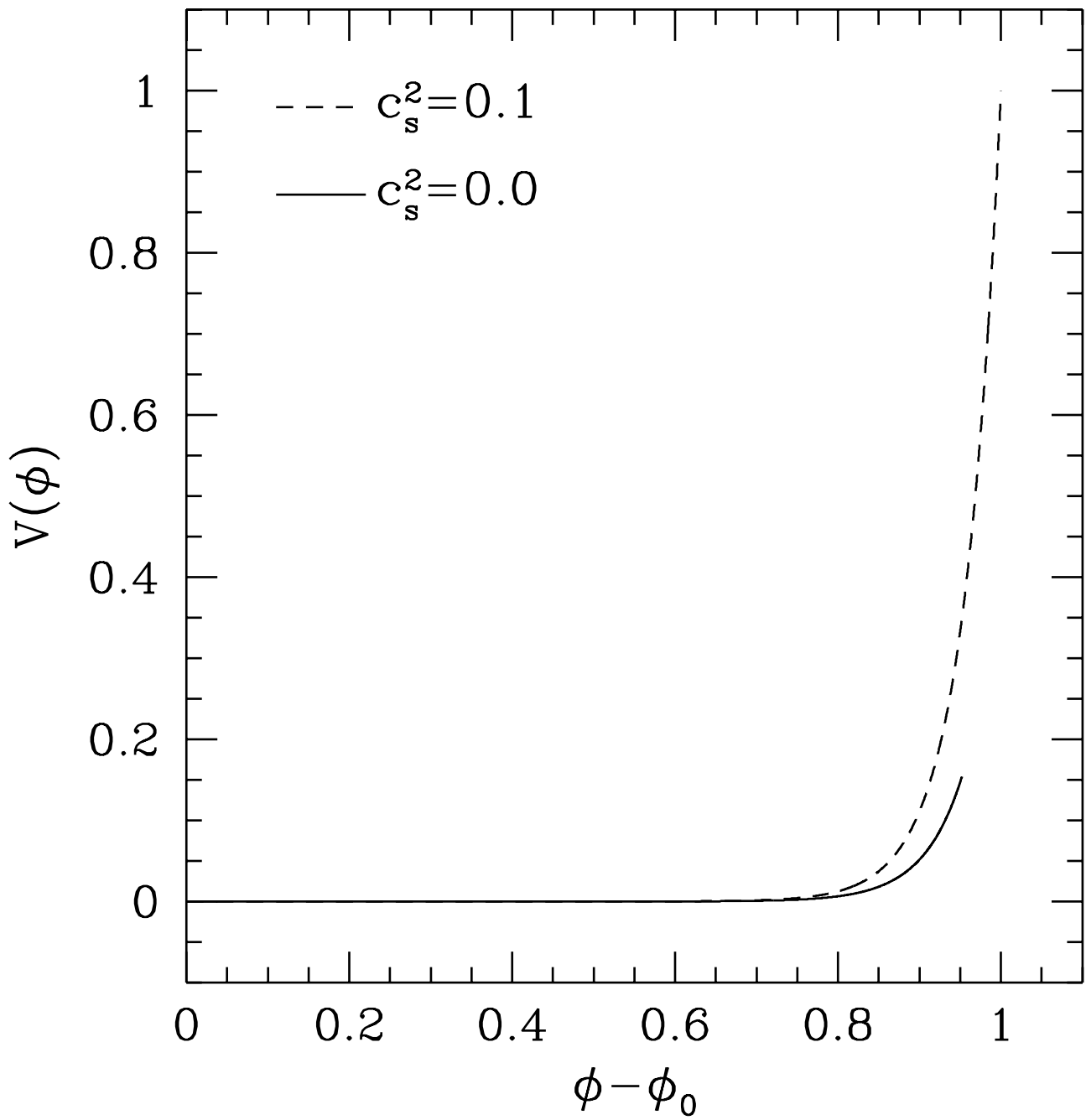
{\bf Figure 14: }
The effective scalar field potential for x-matter model with
$c_s^2=0,0.1$ and $w_0=-0.8$. Scales are arbitrary.

\label{lastpage}

\end{document}